\documentclass[12pt,namedreference]{kluwer}
\usepackage{amssymb}

\usepackage[]{graphicx}
\usepackage{makeidx}

\makeindex

\begin{document}

\bigskip \bigskip %\begin{article}
\begin{opening}         
\title{Mesogranulation and Turbulence in Photospheric Flows} 
\author{J.~K. \surname{Lawrence}}
\author{A.~C. \surname{Cadavid}}  
\runningauthor{J.K. Lawrence, A.C. Cadavid and A. Ruzmaikin}
\runningtitle{Mesogranulation and Turbulence in Photospheric Flows}
\institute{Department of Physics and Astronomy,
California State University Northridge
Northridge, CA 91330-8268, U.S.A. (e-mail: jlawrenc@galileo.csun.edu)}
\author{A. \surname{Ruzmaikin}}
\institute{Jet Propulsion Laboratory, California Institute of Technology, 
4800 Oak Grove Drive, Pasadena, CA 91109, U.S.A.}
\date{February 27, 2001}

\begin{abstract}
Below the scale of supergranules we find that cellular flows are present in the solar 
photosphere at two distinct size scales, approximately $2$ Mm and $4$ Mm, with distinct 
characteristic times. Simultaneously present in the flow is a non-cellular 
component, with turbulent scaling properties and containing $30 \%$ of the flow energy. 
These results are obtained by means of wavelet 
spectral analysis and modeling of vertical photospheric motions in a 2-hour sequence 
of 120 SOHO/MDI, high resolution, Doppler images near disk center. The wavelets permit 
detection of specific local flow patterns corresponding to convection cells.
\end{abstract}

\end{opening}           

\section{Introduction}

The observed velocity field of the solar photosphere falls into two distinct
categories. One is acoustic ``p-waves'' believed to be excited by convective
motions \cite{cox91}. The p-waves inhabit distinctive ridge-like features in
a space-time Fourier spectrum (``$k-\omega $ diagram'') of the motions and
can be isolated numerically in a sequence of Doppler images. The other
category is flows driven by solar convection. The interaction of convection
with the Sun's rotation produces large-scale flows, such as differential
rotation and meridional circulation. These, and giant cells, will not be
considered here. The remaining smaller scale flows also contain structure.
Most apparent are convection cells with a particular topology: an isolated
central upflow surrounded by a multiply connected boundary of downflow \cite
{simon89}. These cells favor certain sizes, particularly the granular ($\sim
1-2$ Mm) and supergranular ($\sim 20-40$ Mm) scales. More controversial are
the intermediate mesogranular scale ($\sim 4-7$ Mm) and the possible
presence of flows that do not have the standard cellular form, that show a
continuum of scales between granules and supergranules, and may represent
turbulence. In this paper we will argue for the presence in the photosphere
of both mesogranular cells and turbulent flows.

Since their discovery \cite{nov81} the sizes and lifetimes of mesogranular
cells have proven difficult to determine accurately. For example,
correlation tracking of a $45.5$ hour sequence of SOHO/MDI high resolution
Doppler images \cite{shine00} find mesogranule scales of $5-10$ arcsec ($4-7$
Mm) with lifetimes of a few hours. Similar size scales were estimated by 
Deubner (1989) on the basis of coherent velocity and intensity fluctuations.
Other workers \cite{ueno98} have
estimated a characteristic size for mesogranules as large as $18$ arcsec ($%
13 $ Mm) and lifetimes peaking between $30-40$ min. For a listing of other
studies see Rieutord, et al. (2000).

Analyses of the photospheric flows in Fourier basis functions or spherical
harmonics show broad spectra, with power distributed across all scales
between granules and supergranules. For example, Ginet and Simon (1992) show
that the Fourier spectrum requires the presence of flow structures at scales 
$\sim 4-7$ Mm, but they cannot establish that the flow producing this power
is cellular or distinct from granules or supergranules. Expansion of both
high and low resolution SOHO/MDI Doppler images in spherical harmonics \cite
{hath00} produces very broad spectra with two clear peaks, but with no sign
of an intermediate peak signifying mesogranules. It is instead suggested
that the spectrum can be decomposed into just two, overlapping, granular and
supergranular distributions. The granular distribution would extend over
scales from below resolution up to more than $100$ Mm, with a broad peak
centered near $3.5$ Mm. The supergranular distribution would extend from $10$
to $300$ Mm in scale with a peak at $\sim 40$ Mm.

Rieutord, et al. (2000) have suggested that different size scales for
mesogranules are reported because of different temporal averaging of their
respective data sets. Photospheric flows on scales larger than granules may
be turbulent and hence composed of flow eddies with sizes obeying a
continuous power law distribution. It also is characteristic of steady
turbulence that the decorrelation times of eddies increase with their size,
also in a power law. Thus, when an observer averages images over some
particular period of time, small eddies with a shorter decorrelation time
will be suppressed and a characteristic scale will be selected, that might
be misinterpreted as mesogranulation.

Here we attempt to resolve this apparent conflict between spectral and
correlation tracking analyses by use of wavelet transforms. Wavelets relate
the photospheric flow fields to localized basis functions. This has the
particular advantage of permitting us to look for particular local
configurations in the overall flow field. We will focus on patterns with the
particular convection cell topology, that is, isolated areas of upflow (blue
shift) bordered by downflow (red shifts). We define this specific pattern to
be ``cellular'' flow. In random flows, this pattern and its reverse should
occur with equal probability; in a field of convection cells, this pattern
will outweigh its reverse. We determine those size scales for which the
correct ``cellular'' signature exceeds its reverse. In the data set we use,
high resolution SOHO/MDI Doppler images at disk center, we will find two,
distinct scales of such cellular flows, centered at about $2$ Mm and $4$ Mm (%
$3$ and $6$ arcsec), and we associate these with granules and mesogranules,
respectively. Although the existence of supergranules is well established in
horizontal flows, they do not appear readily in our analysis of essentially
vertical flows at disk center. However, the presence of spectral power at
scales up to $\sim 50$ Mm, with parity between cellular patterns and their
reverse, implies that significant non-cellular flows are present. Previous
studies by the authors have shown that such structures decorrelate in times
which depend on their scales in a power law relationship. In the present
case, modeling will further indicate power law spatial correlations for
these flows. The scaling properties imply that the non-cellular flows are
turbulent.

It thus appears that standard spectral techniques have difficulty resolving
mesogranules for three basic reasons: (1) the mesogranules are near in scale
to granules and weaker in velocity and (2) they are hidden by overlying
turbulence because (3) global basis functions, such as Fourier waves or
spherical harmonics, do not allow attention to be paid to the local
topologies that label cellular flows. Consequently, determinations of
mesogranular scale that do not take into account the specific flow patterns
may be contaminated by the non-cellular flows and be artifacts of time
averages as suggested by Rieutord, et al. (2000). Nevertheless, when we do
look at the specific convective cell pattern, the mesogranules are indeed
present, at the small end of the range of previously published sizes.

In Section 2 we describe the data set employed in this paper. In section 3
we give a brief summary of continuous wavelet analysis, and in Section 4
this is applied to observed photospheric flow fields. In Section 5 we
construct a model flow field and compare the generated spectra to the
observed case. Conclusions are presented in Section 6.

\section{Data}

\begin{figure}[tbp]
\tabcapfont
\centerline{\begin{tabular}{c@{\hspace{1pc}}c}
\includegraphics[width=2.2in]{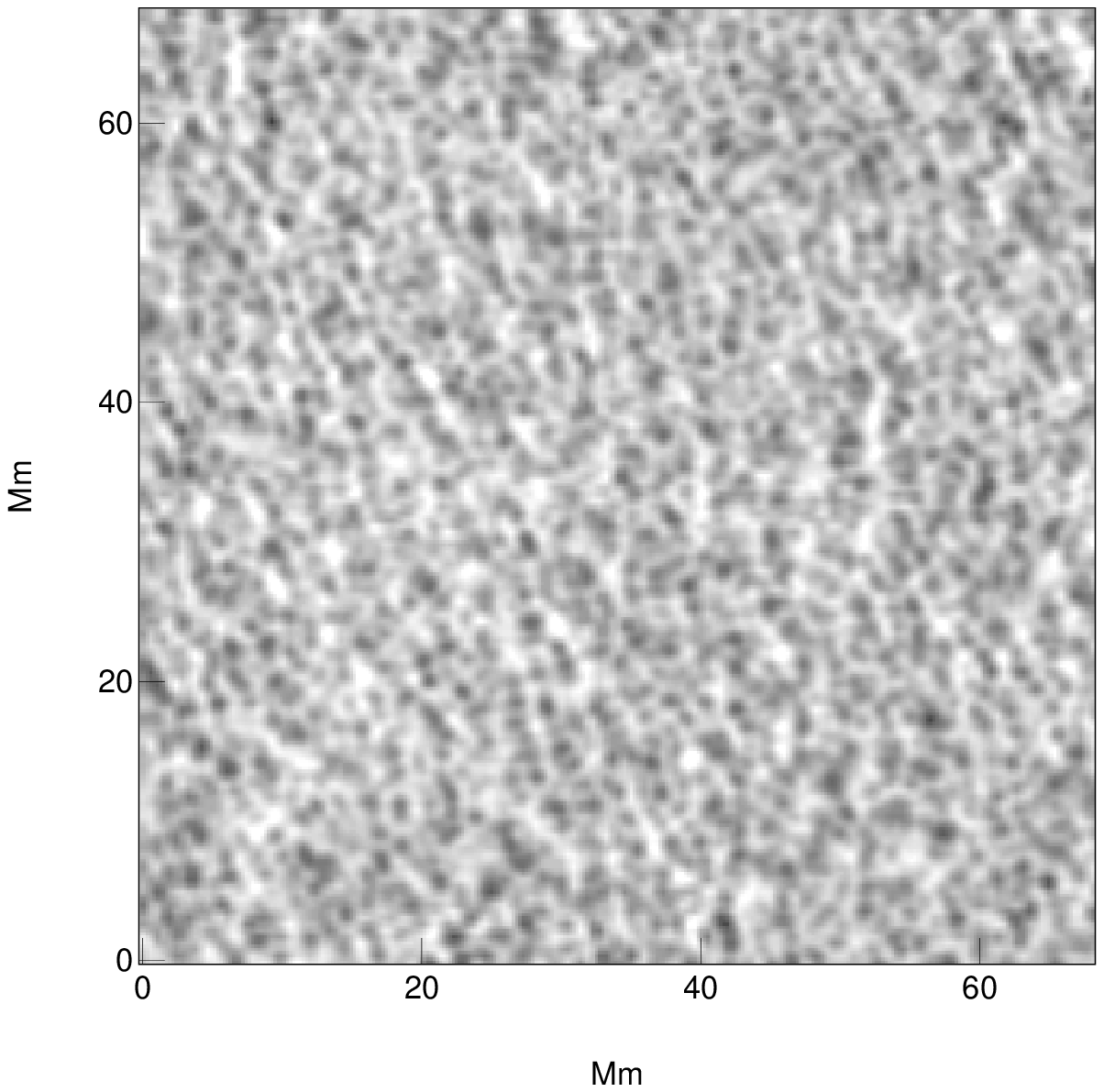} &
\includegraphics[width=2.2in]{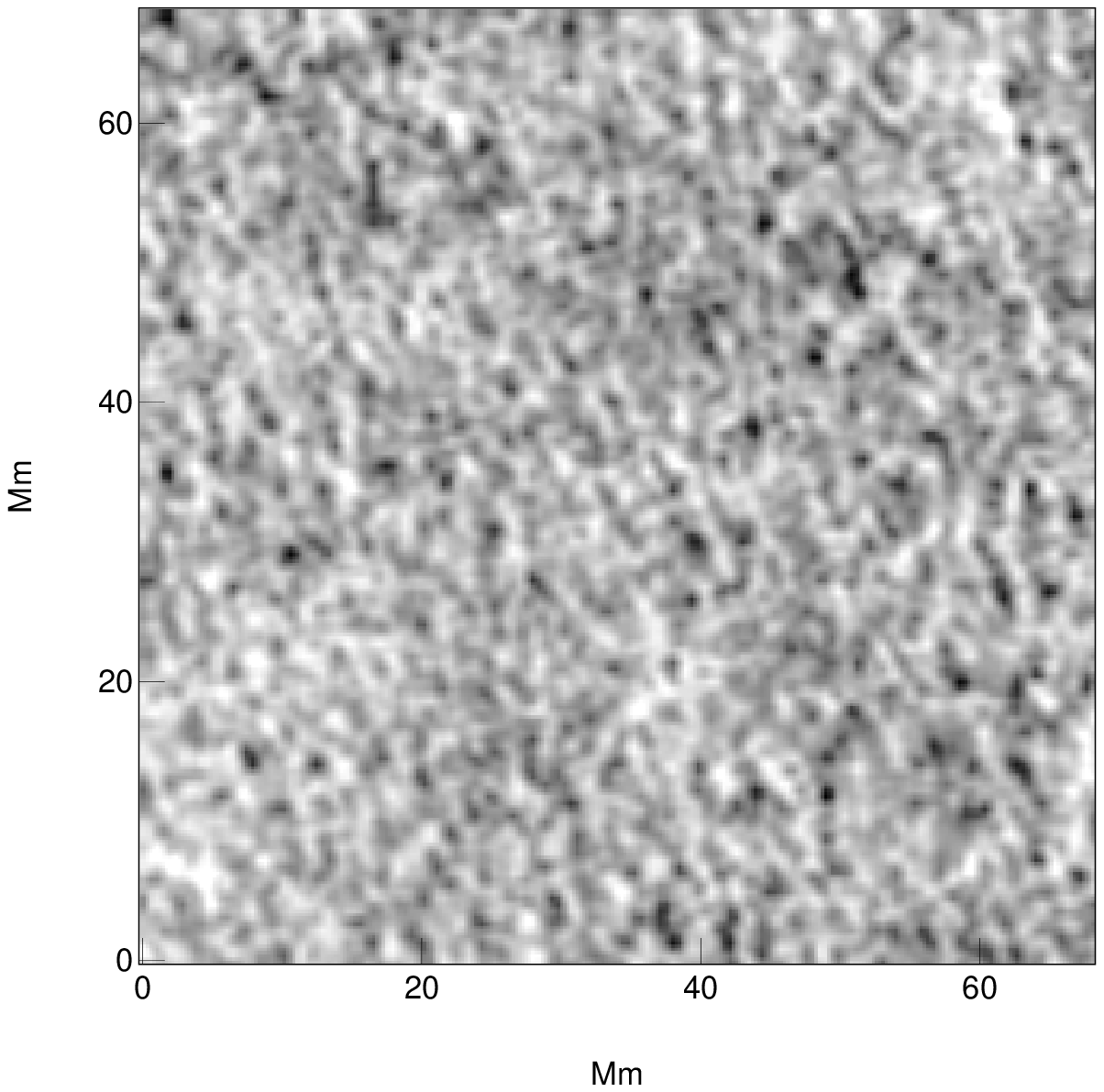} \\
a.~~ Individual Image& b.~~ 120 Min. Average
\end{tabular}}
\caption{a. Single SOHO/MDI, acoustically filtered, Doppler image of
vertical flows near disk center. b. An average of the 120 images spaced at 1
min intervals and spanning two hours in time. In these images red shifted
downflows appear light and blue shifted upflows appear dark.}
\label{twopix}
\end{figure}

The data are SOHO/MDI, high resolution, Doppler images \cite{scher95}. They
consist of a $120$ min sequence of consecutive, $1$ min cadence, $1000$
pixels east-west by $600$ pixels north-south, Doppler images made on 1997
February 5. The pixel scale is $0.625$ arcsec or $0.46$ Mm on the Sun. The
images were defocused to avoid aliasing. To remove p-waves we used a
previously determined optimal acoustic filter \cite{lawr99}: keeping in the $%
k-\omega $ diagram only spatial frequencies $k>5\omega -15$, where $k$ is in
pixels$^{-1}$, and the temporal frequency $\omega $ is in minutes$^{-1}$. So
as to concentrate on vertical photospheric motions we restricted our
attention to a region within $75$ pixels = $34$ Mm of the disk center.
Figure~\ref{twopix} shows an individual Doppler image at the middle of the
sequence, together with an average of the 120 images spanning two hours in
time. Note that the features in the averaged image are larger in scale than
those in the individual image.

For images like those in Figure~\ref{twopix}, we express the vertical
velocity of the photosphere in km s$^{-1}$ as a function of position on the
solar surface: $v(\mathbf{r})$. The convention for the SOHO/MDI Doppler
images is that for red shifted downflows $v>0$; for blue shifted upflows $%
v<0 $.

\section{Wavelet Analysis}

The methods used here are described in more detail in Lawrence, et al
(1999). In two spatial dimensions a continuous wavelet transform $W(s,%
\mathbf{x})$ of a function $v(\mathbf{r})$ is its convolution with scaled
and translated versions of a basis function $\Psi (\mathbf{r})$: 
\begin{equation}
W(s,\mathbf{x})=\frac{1}{s}\int v(\mathbf{r})\Psi (\frac{\mathbf{r-x}}{s}%
)d^{2}\mathbf{r.}
\end{equation}
Note that the transform adds a scale variable $s$ to the position variable $%
\mathbf{x}$. The basis function must be localized in both Fourier space ($%
\mathbf{k}$) and configuration space ($\mathbf{r}$), so that equation (1)
gives a strong response to structure in $v(\mathbf{r})$ of scale $s$ at
position $\mathbf{x}$ and a weak response otherwise.

A variety of basis functions are acceptable. For determining spectra will
use a sixth order function with radial symmetry: 
\begin{equation}
\Psi _{6}(r)=-\nabla ^{2}\nabla ^{2}\nabla ^{2}\exp (-r^{2}/2),
\end{equation}
or in Fourier space 
\begin{equation}
\widetilde{\Psi }_{6}(k)\propto k^{6}\exp (-k^{2}/2).
\end{equation}
This wavelet strongly suppresses both small and large scale (respectively,
large and small $k$) contributions and thus gives good resolution in scale 
\cite{perr95}.

The global wavelet spectrum of an image may be expressed \cite{farge92} as 
\begin{equation}
\mathcal{E}_{W}(s)=\frac{1}{s^{2}}\int \left| W(s,\mathbf{x})\right|
^{2}d^{2}\mathbf{x,}
\end{equation}
where $\left| W(s,\mathbf{x})\right| ^{2}/s^{2}$ is the$\ $energy density as
a function of scale and position. The global energy spectrum is related to
the Fourier transform $\widetilde{v}(\mathbf{k})$ of $v(\mathbf{r})$ \ and
the Fourier energy density $\mathcal{E}_{F}(\mathbf{k})=\left| \widetilde{v}(%
\mathbf{k})\right| ^{2}$ by 
\begin{equation}
\mathcal{E}_{W}(s)=\int \mathcal{E}_{F}(\mathbf{k})\left| \widetilde{\Psi }(s%
\mathbf{k})\right| ^{2}d^{2}\mathbf{k,}
\end{equation}
where $\widetilde{\Psi }$ is the Fourier transform of $\Psi $. Thus $%
\mathcal{E}_{W}(s)$ is the Fourier energy density, smoothed by the Fourier
spectrum of the wavelet basis function at each scale. The information it
carries is similar to that carried by the Fourier density, but with lower
resolution in scale or spatial frequency.

\section{Data Analysis}

\subsection{Fourier Spectra}

\begin{figure}[tbp]
\centerline{\includegraphics[width=4in]{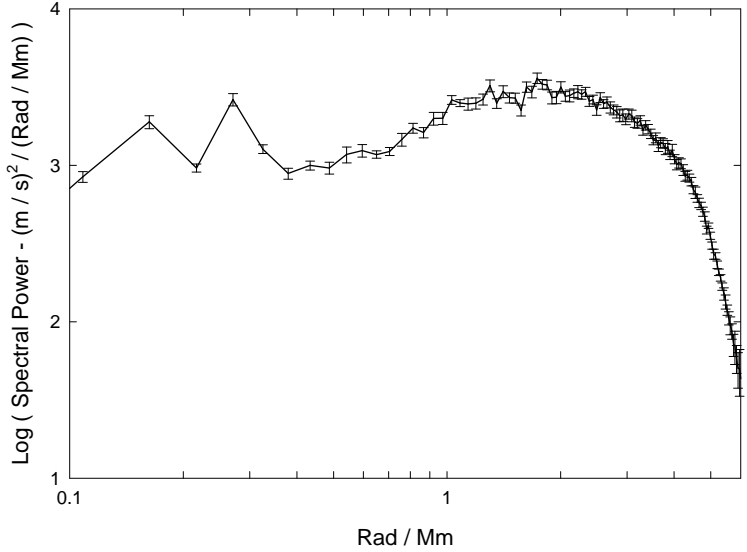}}
\caption{$E_{F}(k)$ Fourier power spectrum. The error bars are derived from
the average over individual spectra.}
\label{foucalib}
\end{figure}

The total power in an image may be written 
\begin{equation}
E=\int d^{2}k\left| \widetilde{v}(\mathbf{k})\right| ^{2}=\int_{0}^{\infty
}dkk\int_{0}^{2\pi }d\phi \mathcal{E}_{F}(k)
\end{equation}
Here $k=\left| \mathbf{k}\right| $ and $\phi $ are polar coordinates in
Fourier space. For the 2-dimensional images the Fourier power spectrum $%
E_{F}(k)$, or power per unit wave number, is 
\begin{equation}
E_{F}(k)=dE/dk=k\int_{0}^{2\pi }d\phi \mathcal{E}_{F}(k),
\end{equation}
where we assume statistical isotropy. For a white noise image $%
E_{F}(k)\propto k$, and for Kolmogorov turbulence $E_{F}(k)\propto k^{-2/3}$%
.\ Here we make use only of squares of side $256$ pixels = $116$ Mm centered
on disk center. Figure~\ref{foucalib} depicts the average of the Fourier
spectra of the image centers. The $1\sigma $ error bars are derived from the
scatter in the spectra of invividual images spaced at 10 minute intervals.
This allows some randomization and statistical independence in the small
scales.

We will use the Fourier spectrum to calibrate the spectral power of the
wavelet spectra.

\subsection{Wavelet Spectra}

We compute the convolutions of the images with the $\Psi _{6}$ basis
function according to equation (1) and the wavelet energy density according
to equation (4). The spectra will be plotted as functions of spatial
frequency $k=2\pi /s$, and transformed to the format of a Fourier spectrum,
that is, as power per unit wavenumber $k$. The resulting wavelet power
spectrum, as in equation (7), is 
\begin{equation}
E_{W}(k)=k\int_{0}^{2\pi }\mathcal{E}_{W}(k)d\phi ,
\end{equation}
although calculated with wavelets, is directly comparable to the Fourier
spectrum $E_{F}(k)$.

Two kinds of wavelet spectra are computed. Since the wavelet basis function
is a positive central peak with radial ripples around it, the wavelet
transform $W(s,\mathbf{x})$ of a convection cell, with a central $v<0$
upflow and peripheral $v>0$ downflow, will be negative at the cell's scale $%
s $ and position $\mathbf{x}$. In the case of a random flow field, $W(s,%
\mathbf{x})$ at a given scale and location will be positive or negative with
equal probability. If convection cells also are present in the flow there
will be an excess of negative values around the corresponding locations and
scales.This local feature of the wavelet transforms is what we use to look
for mesogranulation.

\begin{figure}[tbp]
\centerline{\includegraphics[width=4in]{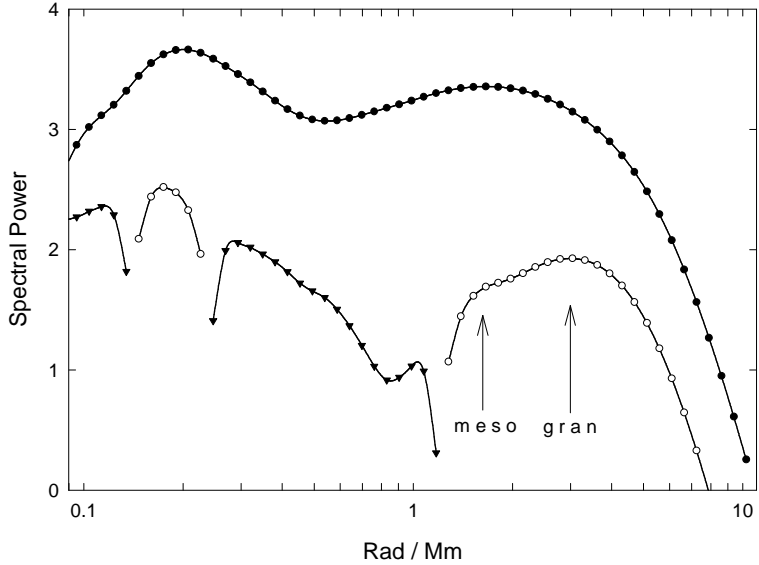}}
\caption{Filled circles: average of the wavelet power spectra $E_{W} (k) $
of the 120 photospheric images. Open circles: positive part of the average
of the cellular spectra $E_{C} (k)$ of the images. Filled triangles: average
of the negative part of the cellular spectra. This indicates the relative
significance of positive excursions of $E_{W}$. The power is calibrated to
the corresponding Fourier spectrum in Figure 2.}
\label{wavela}
\end{figure}

We compute, within $34$ Mm of the disk center, the wavelet spectrum $%
E_{W}^{+}(k)$ using only the positive values of $W(s,\mathbf{x})$. Likewise,
we compute the spectrum $E_{W}^{-}(k)$ including only the negative values.
The $E_{W}^{+}(k)$ spectrum will include essentially no contributions from
convective ``cells'' (which correspond to $W<0$), but only non-cellular
contributions. The $E_{W}^{-}(k)$ spectrum will include both cellular and
non-cellular contributions. Because of the locality property of the
wavelets, the sum $E_{W}(k)=E_{W}^{-}(k)+E_{W}^{+}(k)$ is just the full
wavelet spectrum and is a smoothed version of the Fourier spectrum as
described above. The difference of the two spectra $%
E_{C}(k)=E_{W}^{-}(k)-E_{W}^{+}(k)$ will characterize the cell scales.
Recall that $E_{W}^{-}(k)$ contains all the cellular contributions. The
non-cellular ``background'' is, at least statistically, subtracted off with $%
E_{W}^{+}(k)$. Those scales $s=2\pi /k$ for which $E_{C}(k)$ is
significantly positive will be those scales at which convection cells are
present in the photospheric images. The average of the $E_{W}(k)$ spectra of
the 120 images and the corresponding average cellular $E_{C}(k)$ spectrum
are shown in Figure~\ref{wavela}.

Figure~\ref{wavela} shows the full $E_{W}(k)$ spectrum as filled circles.
Shown as open circles are the parts of the cellular spectrum for which $%
E_{C}(k)>0$. Shown as filled triangles are the absolute values of those
parts of the cellular spectrum for which $E_{C}(k)<0$. These last parts
characterize the non-cellular, background flows. Their negative slope
indicates strong large scale correlations in the stochastic background. The $%
E_{C}>0$ feature near $k=0.2$ Rad Mm$^{-1}$, though suggestively near $30$
Mm in size, is not sufficiently strong to represent supergranules, but
represents non-cellular, large scale correlation features. This will be
discussed further below, when we model the flows. The cellular spectrum in
Figure~\ref{wavela} does, however, indicate the presence of two
characteristic spatial scales: at $\sim 2$ Mm corresponding to granules and
at $\sim 4$ Mm representing mesogranules.

\subsection{Decorrelation Times}

\begin{figure}[tbp]
\centerline{\includegraphics[width=4in]{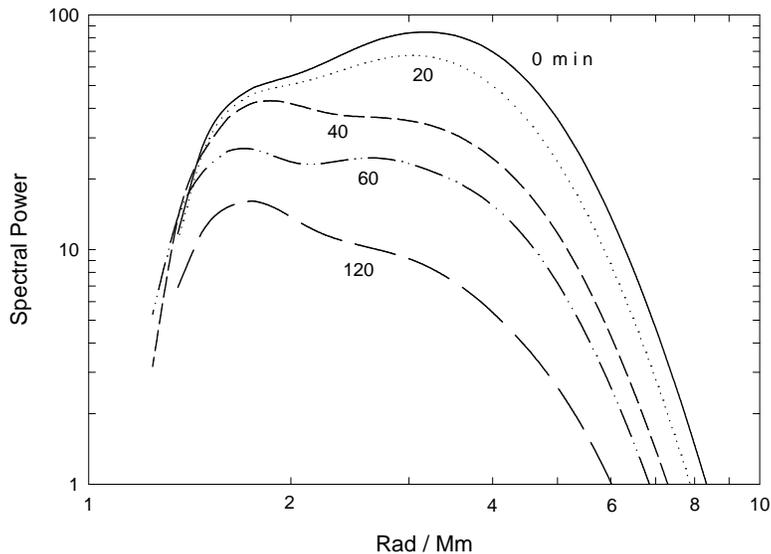}}
\caption{Solid line: average of the cellular spectra of each of the 120
Doppler images. Dotted line: Average of the cellular spectra of six image
averages spanning $20$ min each. Short-dashed line: Average of spectra of
three image averages spanning $40$ min each. Dot-dash line: Average of
spectra of three overlapping image averages spanning $60$ min each.
Long-dashed line: spectrum of an average over all 120 images spanning $120$
min.}
\label{timedec}
\end{figure}

Figure~\ref{timedec} shows as a solid line a magnified view of the granular
and mesogranular cellular $E_{C}(k)$ spectrum. Recall that this is taken
from an average over the individual spectra of all 120 of the Doppler
images. It thus characterizes the stationary flow as seen at a typical point
in time. Shown in Figure~\ref{timedec} as the long dashed line is an
analogous spectrum made by first averaging the 120 images and then
calculating the cellular spectrum. Because this average spans $2$ hr in
time, features with coherence times $\lesssim 2$ hr are blurred, and the
spectral power is accordingly reduced. The granular feature at $k\approx
3.14 $ Rad Mm$^{-1}$ ($s\approx 2$ Mm) has decayed by an order of magnitude,
much more than the mesogranular feature at $k\approx 1.57$ Rad Mm$^{-1}$ ($%
s\approx 4$ Mm). In fact, the mesogranular peak now dominates the granular
one.

Also shown in Figure~\ref{timedec} are plots of the average of six spectra
made from $20$ min averages of images, the average of three spectra made
from $40$ min averages of images, and the average of three spectra made from
(overlapping) $60$ min image averages. These allow us to trace the
evolutionary timescales of the cellular features. The granular feature is
already decayed in the $20$ min averages and has dropped by a factor $\sim 3$
in the $40$ min averages. The mesogranular feature has not decayed
essentially at all in the $40$ min averages and is larger than the granular
feature. By the $60$ min image average the mesogranules are evolving too,
and their peak is decreasing. Thus we see that, in addition to distinct
spatial scales, there are distinct, non-overlapping, coherence times for the
granules (less than $20$ min) and the mesogranules (more than $40$ min).

The essential difference between this result and the suggestion of Rieutord,
et al. (2000) is that averaging of images over various time periods does not
select from a continuum of \textit{cellular} scales, but rather from two
scales. For short averaging times the granular scale $2$ Mm is selected. For
longer times the mesogranular scale $4$ Mm is selected.

\section{Modelled Data}

To extend the interpretation of the observational results, we construct a
model photospheric velocity field. This is intended to apply to the
velocities at a typical moment in time and includes no evolution. We
represent a model ``cell'' by the negative of a ``Mexican hat'' wavelet
basis function in the form 
\begin{equation}
\Psi _{M}(r)=V(1-r^{2}/2a^{2})\exp (-r^{2}/2a^{2})
\end{equation}
This models a central blue shifted feature with characteristic radius $a$
surrounded by a red shifted ring with maximum downflow at radius $2a$. We
take the diameter $4a$ to be the ``size'' of a cell with scale parameter $a$%
. On an array of $1024$ pixels $\times $ $1024$ pixels we randomly locate
with uniform probability $32,768$ ``granular'' wavelets with scale $a=3$
pixels and central vertical velocity $V=110$ ms$^{-1}$ and $3,640$
``mesogranular'' wavelets with $a=6$ pixels and $V=45$ ms$^{-1}$. As we will
see, this models well the granular and mesogranular parts of the cellular
spectrum if we choose $a=0.5$ Mm. The numbers of model granules and
mesogranules included are based on the numbers that would be contained in
square arrays with separations $3a-4a$.

To model the observed spectral power at scales $\gtrsim 4$ Mm ($k\lesssim
1.57$ Rad Mm$^{-1}$) requires a non-cellular flow component with a
continuous range of scales. To model the correct slope for the spectra it is
necessary to use ``colored'' noise with increased spatial correlations at
large scales. We first generate a Gaussian white noise image. Then we
transform to Fourier space and multiply the transform by $k^{-\gamma }$ with
cutoffs in scale $s=2\pi /k<$ $4$ pixels and $s>$ $256$ pixels. The negative
exponent emphasizes larger scales. The spectrum of this colored noise has a
power law form $E(k)\propto k^{1-2\gamma }$. The best fit to solar data is
given by a value $\gamma =0.6$, giving the spectrum $E(k)\propto k^{-0.2}$.
To get the image noise we transform back to configuration space. The
standard deviation of the colored noise used in the model was $42$ ms$^{-1}$.

\begin{figure}[tbp]
\tabcapfont
\centerline{\begin{tabular}{c@{\hspace{1pc}}c}
\includegraphics[width=2.2in]{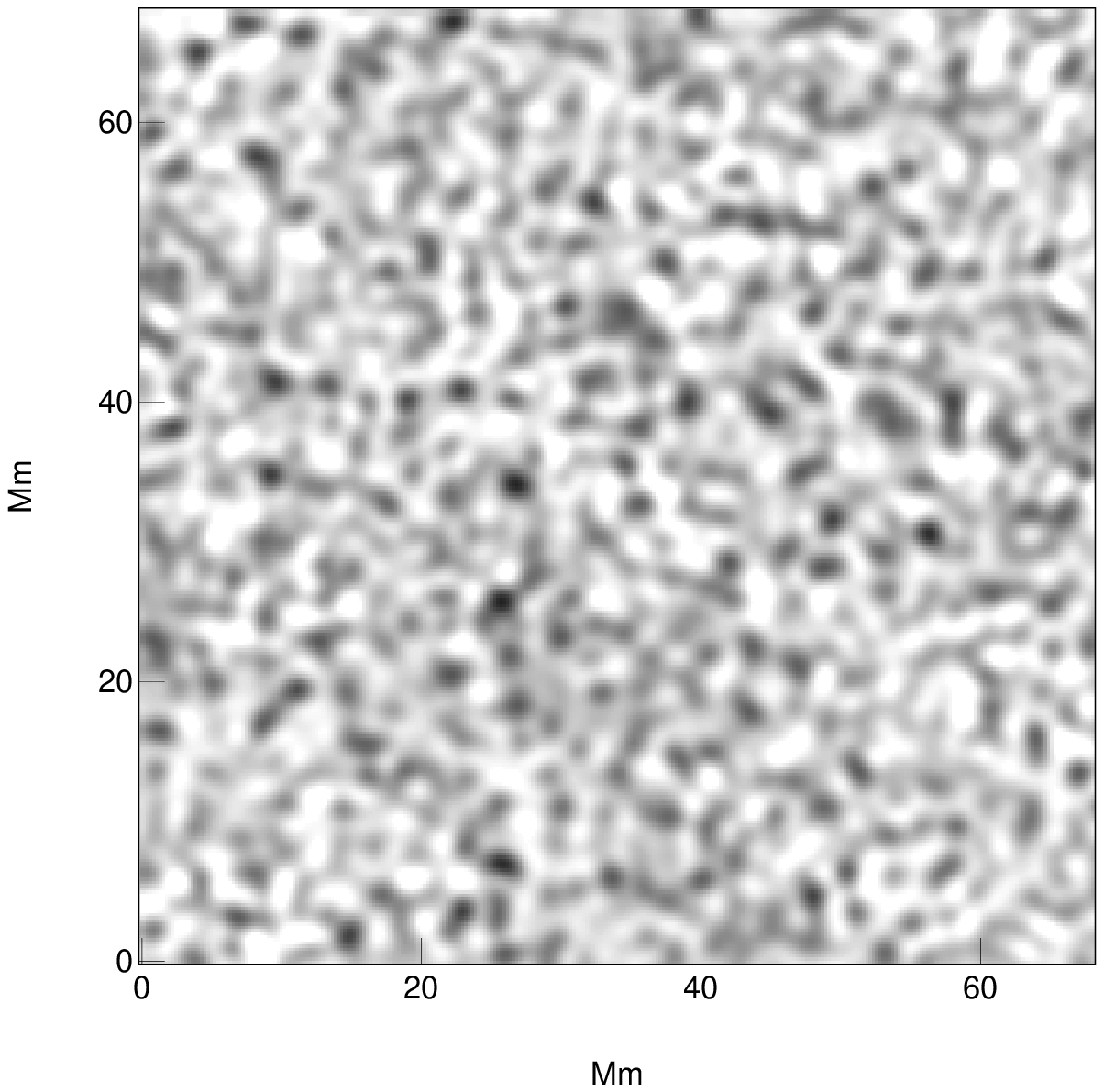} &
\includegraphics[width=2.2in]{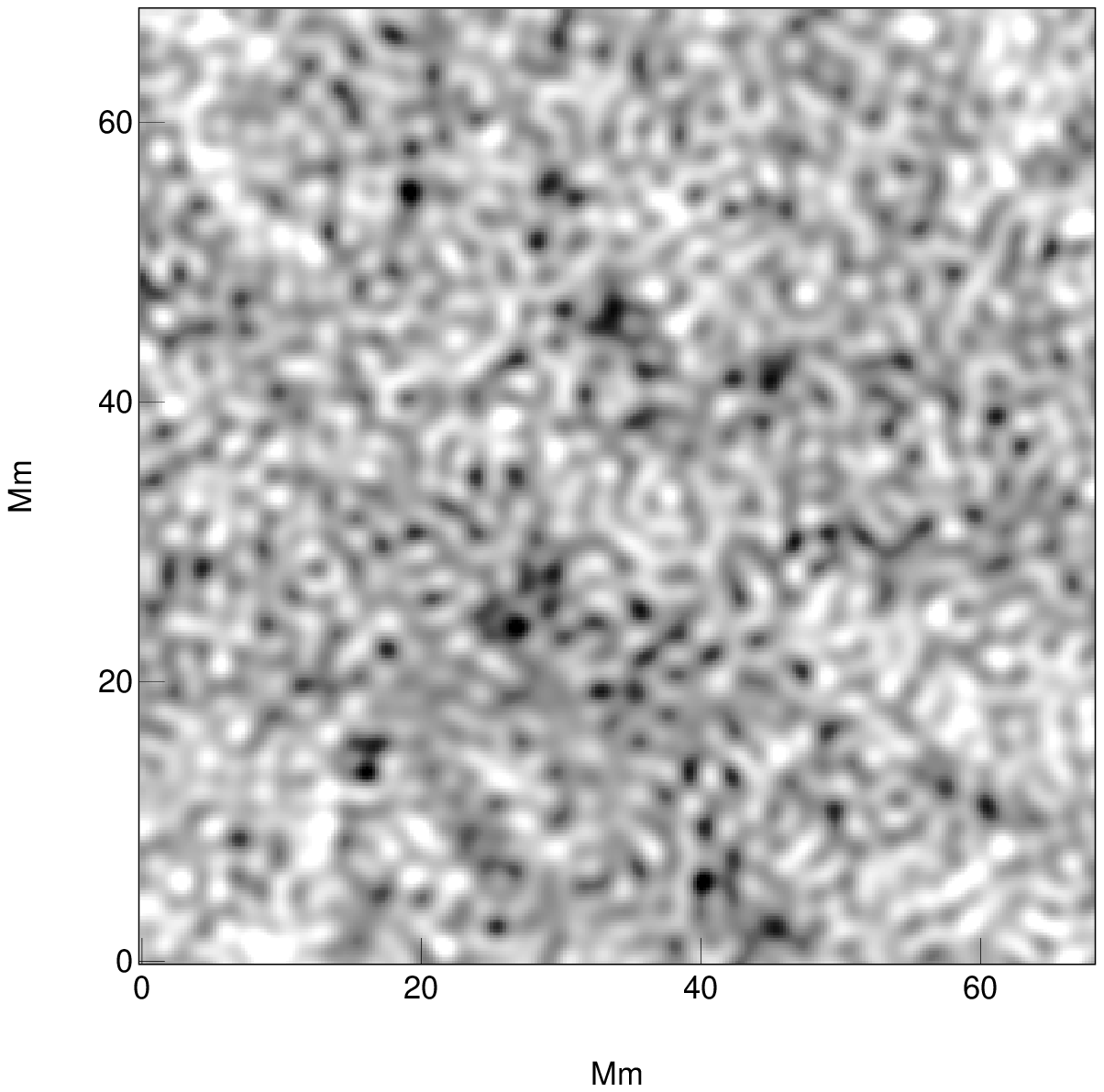} \\
a.~~ Modeled Image& b.~~ Colored Noise
\end{tabular}}
\caption{a. Image of artificial flow field constructed from two discrete
cell sizes and spatial colored noise. b. The colored noise used in (a). Note
the long range spatial correlations.}
\label{modpix}
\end{figure}

The cellular and stochastic velocities quoted here were obtained by
calibrating the overall image variance in Figure \ref{modpix}a to that of
Figure \ref{twopix}a. The variance of the cellular component of the model
image is $4800$ m$^{2}$s$^{-2}$, and that of the stochastic component is $%
1800$ m$^{2}$s$^{-2}$. This implies that the stochastic component contains $%
\sim 30\%$ of the variance and hence kinetic energy of the model flow.

An artificial cell-plus-noise velocity field created in this way is
illustrated in Figure~\ref{modpix}a. An image of the colored noise alone is
shown in Figure~\ref{modpix}b.

\begin{figure}[tbp]
\centerline{\includegraphics[width=4in]{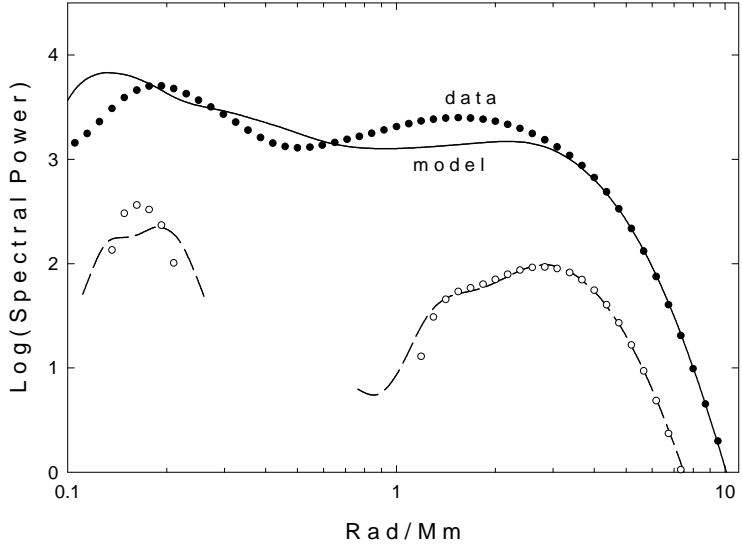}}
\caption{Filled circles: Average of the full $E_{W}(k)$ wavelet spectra of
all 120 images. Open circles: Average of the positive parts of the cellular $%
E_{C}(k)$ spectra over all 120 images. Solid Line: $E_{W}(k)$ spectrum of
the modelled image of Figure 5a. Dashed line: Positive parts of the $%
E_{C}(k) $ spectrum of the modelled image. Frequency units have been
adjusted to Rad Mm$^{-1}$ by calibration with the observed spectra.}
\label{model}
\end{figure}

The wavelet spectra generated by the modelled velocity field in Figure~\ref
{modpix}a are shown in Figure~\ref{model} in comparison to the solar data.
The filled circles indicate $E_{W}(k)$ for the solar data, and the open
circles the positive part of $E_{C}(k)$ for the solar data. The solid line
is $E_{W}(k)$ for the modelled data, and the dashed lines give $E_{C}(k)$
for the modelled data. The slopes for the full $E_{W}(k)$ spectra are in
general agreement at all scales.

The observed and modelled cellular spectra of the granules and mesogranules
agree very well for wave numbers $1$Rad Mm$^{-1}<k<10$ Rad Mm$^{-1}$. The
resolution of $E_{C}(k)$\ is such that the granular and mesogranular peaks
overlap. We know, however, by construction, that the granular and
mesogranular scales in the modelled image are discrete. It follows that
these scales in the solar data may well also be discrete. Attempts to model
the observed cellular spectrum with a continuum of cell sizes between $2$ Mm
and $4$ Mm were unsuccessful.

The particular realization of the modelled data shown here shows a structure
near $k\approx 0.2$ Rad Mm$^{-1}$ or scale $\sim 30$ Mm. Again by
construction, the modelled image contains no supergranules. This structure
must therefore represent large scale correlations in the stochastic
component of the modelled flow. Accordingly, we interpret the comparable
feature in the observed spectrum as a large scale correlation feature of the
non cellular-flow, rather than as supergranular. Recall that the spectrum of
the background ``noise'' has slope $-0.2$.

\begin{figure}[tbp]
\centerline{\includegraphics[width=4in]{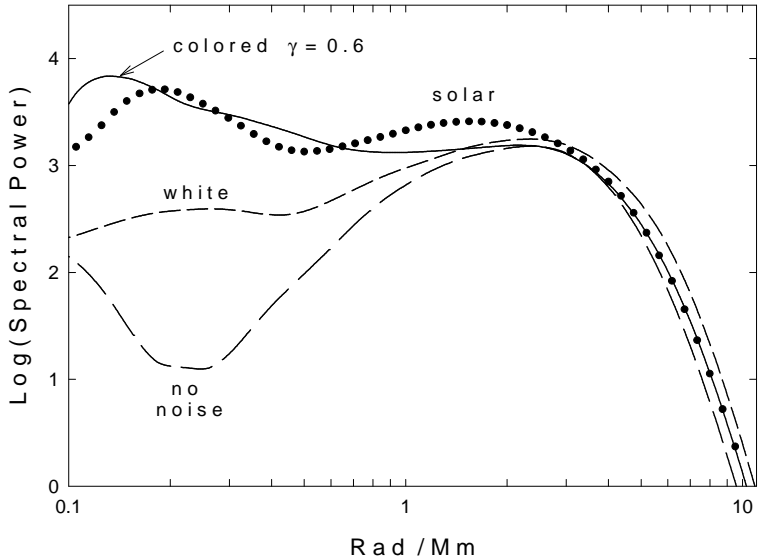}}
\caption{Filled circles: Average of the full $E_{W}(k)$ wavelet spectra of
the 120 solar images. Solid line: Full $E_{W}(k)$ spectrum of the modelled
image of Figure 5a made with a colored noise stochastic component.
Short-dashed line: $E_{C}(k)$ spectrum for the same modelled image, but with
a white noise stochastic component. Long-dashed line: $E_{W}(k)$ spectrum of
the same modelled image, but with no stochastic component. Frequency units
have been adjusted to Rad Mm$^{-1}$ by calibration with the observed
spectra. }
\label{noismod}
\end{figure}

Figure~\ref{noismod} shows the same full wavelet spectra of the observed and
modelled solar flows that are shown in Figure~\ref{model}. Also shown are
other modelled spectra constructed with no added stochastic component (long
dashed line) and with a white noise ($\gamma =0$) stochastic component
(short dashed line). Neither of the last two is able to approximate the
observed spectrum at large scales ($k\lesssim 1$ Rad Mm$^{-1}$).

\section{Discussion and Conclusions}

The cellular spectra of the SOHO/MDI Doppler images cannot be explained
without two discrete cellular components: granules with size scale $\sim 2$
Mm and coherence time less than $20$ min and mesogranules with size scale $%
\sim 4$ Mm and coherence time greater than $40$ min. There is no evidence
that the granules and mesogranules overlap significantly in scale. We note
that the convective cells visible in Figure~\ref{twopix}b are noticeably
larger than in Figure~\ref{twopix}a and, based on our analysis, depict
mesogranules.

The photospheric velocity spectrum can be reproduced in all its essential
aspects by adding to the modelled cellular flows a stochastic component with
power law spatial correlations, resembling those of turbulence, and
containing $\sim 30\%$ of the flow energy. In earlier work \cite{cad98} we
found that the stochastic component of vertical flows in high resolution
SOHO/MDI Doppler images displayed decorrelation times that scaled as a power
law with spatial frequency. These two kinds of scaling indicate fluid
turbulence. We have found a best fit spatial spectral index $\gamma =0.6$
which is similar to the $\gamma \approx 0.83$ which would signify Kolmogorov
turbulence. Previous work by the authors \cite{cad98} has indicated $\gamma
\approx 0.83$ for horizontal flows from SOHO full disk data for scales
greater than $16$ Mm\ ($k<0.4$ Rad Mm$^{-1}$). The horizontal flows also
show temporal decorrelation scaling over scales $3.6$ Mm $\leq s\leq $ $120$
Mm. Similar results were found still earlier \cite{ruz96} for a sequence of
San Fernando Observatory Doppler images.

With a photospheric Reynolds number $\sim 10^{8}$ the presence of turbulence
should be no surprise. The co-existence of stochastic flows and coherent
structures determined by boundary conditions is a hallmark of turbulence in
nature \cite{holm96}. We thus suggest that, after careful filtering for
p-waves, the photospheric flows indeed fall into both of these categories.
The cellular flows appear to be present at discrete scales which must be
determined in some way by convective boundary conditions. The textbook
explanation in terms of the recombination depths of hydrogen at 1 Mm and of
helium at 5 Mm and 15 Mm (for supergranules) \cite{simon68} remains viable.

\acknowledgements
SOHO is a project of international cooperation between ESA and NASA. This
work was supported in part by NSF grant ATM-9987305.

%\end{article}

\end{document}